\documentclass[epjmod,nopacs,a4paper,10pt]{svjourmod}

\usepackage[english]{babel}
\usepackage{amsmath}
\usepackage{amssymb}
\usepackage{graphicx}
\usepackage{styles/units}
\usepackage{epsfig} 
\begin{document}
\hugehead
\title{\boldmath A Measurement of the $\psi^\prime$ to $J/\psi$ 
Production Ratio
in 920 GeV Proton-Nucleus Interactions}

\author{
I.~Abt\inst{23}\and
M.~Adams\inst{10}\and
M.~Agari\inst{13}\and
H.~Albrecht\inst{12}\and
A.~Aleksandrov\inst{29}\and
V.~Amaral\inst{8}\and
A.~Amorim\inst{8}\and
S.~J.~Aplin\inst{12}\and
V.~Aushev\inst{16}\and
Y.~Bagaturia\inst{12\and36}
% ...
}

\mail{Alexander.Spiridonov@desy.de}

\institute{
$^{1}${\it NIKHEF, 1009 DB Amsterdam, The Netherlands~$^{a}$} \\
$^{2}${\it Department ECM, Faculty of Physics, University of Barcelona, E-08028 Barcelona, Spain~$^{b}$} \\
$^{3}${\it Institute for High Energy Physics, Beijing 100039, P.R. China} \\
$^{4}${\it Institute of Engineering Physics, Tsinghua University, Beijing 100084, P.R. China} \\
$^{5}${\it Institut f\"ur Physik, Humboldt-Universit\"at zu Berlin, D-12489 Berlin, Germany~$^{c,d}$} \\
$^{6}${\it Dipartimento di Fisica dell' Universit\`{a} di Bologna and INFN Sezione di Bologna, I-40126 Bologna, Italy} \\
$^{7}${\it Department of Physics, University of Cincinnati, Cincinnati, Ohio 45221, USA~$^{e}$} \\
$^{8}${\it LIP Coimbra, P-3004-516 Coimbra,  Portugal~$^{f}$} \\
$^{9}${\it Niels Bohr Institutet, DK 2100 Copenhagen, Denmark~$^{g}$} \\
$^{10}${\it Institut f\"ur Physik, Universit\"at Dortmund, D-44221 Dortmund, Germany~$^{d}$} \\
$^{11}${\it Joint Institute for Nuclear Research Dubna, 141980 Dubna, Moscow region, Russia} \\
$^{12}${\it DESY, D-22603 Hamburg, Germany} \\
$^{13}${\it Max-Planck-Institut f\"ur Kernphysik, D-69117 Heidelberg, Germany~$^{d}$} \\
$^{14}${\it Physikalisches Institut, Universit\"at Heidelberg, D-69120 Heidelberg, Germany~$^{d}$} \\
$^{15}${\it Department of Physics, University of Houston, Houston, TX 77204, USA~$^{e}$} \\
$^{16}${\it Institute for Nuclear Research, Ukrainian Academy of Science, 03680 Kiev, Ukraine~$^{h}$} \\
$^{17}${\it J.~Stefan Institute, 1001 Ljubljana, Slovenia~$^{i}$} \\
$^{18}${\it University of Ljubljana, 1001 Ljubljana, Slovenia} \\
$^{19}${\it University of California, Los Angeles, CA 90024, USA~$^{j}$} \\
$^{20}${\it Lehrstuhl f\"ur Informatik V, Universit\"at Mannheim, D-68131 Mannheim, Germany} \\
$^{21}${\it University of Maribor, 2000 Maribor, Slovenia} \\
$^{22}${\it Institute of Theoretical and Experimental Physics, 117259 Moscow, Russia~$^{k}$} \\
$^{23}${\it Max-Planck-Institut f\"ur Physik, Werner-Heisenberg-Institut, D-80805 M\"unchen, Germany~$^{d}$} \\
$^{24}${\it Dept. of Physics, University of Oslo, N-0316 Oslo, Norway~$^{l}$} \\
$^{25}${\it Fachbereich Physik, Universit\"at Rostock, D-18051 Rostock, Germany~$^{d}$} \\
$^{26}${\it Fachbereich Physik, Universit\"at Siegen, D-57068 Siegen, Germany~$^{d}$} \\
$^{27}${\it Institute for Nuclear Research, INRNE-BAS, Sofia, Bulgaria} \\
$^{28}${\it Universiteit Utrecht/NIKHEF, 3584 CB Utrecht, The Netherlands~$^{a}$} \\
$^{29}${\it DESY, D-15738 Zeuthen, Germany} \\
$^{30}${\it Physik-Institut, Universit\"at Z\"urich, CH-8057 Z\"urich, Switzerland~$^{m}$} \\
$^{31}${\it visitor from Dipartimento di Energetica dell' Universit\`{a} di Firenze and INFN Sezione di Bologna, Italy} \\
$^{32}${\it visitor from P.N.~Lebedev Physical Institute, 117924 Moscow B-333, Russia} \\
$^{33}${\it visitor from Moscow Physical Engineering Institute, 115409 Moscow, Russia} \\
$^{34}${\it visitor from Moscow State University, 119899 Moscow, Russia} \\
$^{35}${\it visitor from Institute for High Energy Physics, Protvino, Russia} \\
$^{36}${\it visitor from High Energy Physics Institute, 380086 Tbilisi, Georgia} \\
$^\dagger${\it deceased}
\vspace{5mm}\\
$^{a}$ supported by the Foundation for Fundamental Research on Matter (FOM), 3502 GA Utrecht, The Netherlands \\
$^{b}$ supported by the CICYT contract AEN99-0483 \\
$^{c}$ supported by the German Research Foundation, Graduate College GRK 271/3 \\
$^{d}$ supported by the Bundesministerium f\"ur Bildung und Forschung, FRG, under contract numbers 05-7BU35I, 05-7DO55P, 05-HB1HRA, 05-HB1KHA, 05-HB1PEA, 05-HB1PSA, 05-HB1VHA, 05-HB9HRA, 05-7HD15I, 05-7MP25I, 05-7SI75I \\
$^{e}$ supported by the U.S. Department of Energy (DOE) \\
$^{f}$ supported by the Portuguese Funda\c c\~ao para a Ci\^encia e Tecnologia under the program POCTI \\
$^{g}$ supported by the Danish Natural Science Research Council \\
$^{h}$ supported by the National Academy of Science and the Ministry of Education and Science of Ukraine \\
$^{i}$ supported by the Ministry of Education, Science and Sport of the Republic of Slovenia under contracts number P1-135 and J1-6584-0106 \\
$^{j}$ supported by the U.S. National Science Foundation Grant PHY-9986703 \\
$^{k}$ supported by the Russian Ministry of Education and Science, grant SS-1722.2003.2, and the BMBF via the Max Planck Research Award \\
$^{l}$ supported by the Norwegian Research Council \\
$^{m}$ supported by the Swiss National Science Foundation
}

%
\abstract{
Ratios of the $\psi^\prime$ 
over the $J/\psi$  production cross sections
in the dilepton channel for C, Ti and W targets have been measured
in 920\,GeV proton-nucleus interactions with the HERA-B detector at the HERA
storage ring. 
The $\psi^\prime$ and $J/\psi$ states were reconstructed 
in both the $\mu^+\mu^-$ and the $e^+e^-$ decay modes. 
The measurements covered the kinematic range
$-0.35 \le x_F \le 0.1$ with transverse momentum 
$p_T \le 4.5\,{\rm GeV}/c$.  
The $\psi^\prime$ to $J/\psi$ production
ratio is almost constant in the covered $x_F$ range and 
shows a slow increase with $p_T$.
The angular dependence of the ratio has been used to measure the
difference of the $\psi^\prime$ and $J/\psi$ polarization.
All results for the muon and electron decay channels
are in good agreement: their ratio,
averaged over all events, is 
$R_{\psi'}(\mu)/R_{\psi'}(e)=1.00\pm0.08\pm0.04$. 
This result constitutes a new, direct experimental constraint on 
the double ratio of branching fractions, 
$(B'(\mu) \cdot B(e))\,/\,(B(\mu) \cdot B'(e))$, of $\psi'$
and $J/\psi$ in the two channels.
%
} 
\maketitle

\section{\boldmath Introduction} 
\label{intro} 
The hadroproduction of charmonium can be
described by 
models\,[1--4]
which rely on measured structure functions to describe partons in the 
initial state, perturbative QCD to calculate the production
of intermediate $c\bar{c}$  states,
and a subsequent non-perturbative hadronization step.
Depending on the model assumptions and scope, this latter step must
account for several effects, including color neutralization by soft gluon
emission, bound state formation and the influence of nuclear matter
(``nuclear suppression''). This in turn leads to the introduction of
free parameters which are tuned to describe existing data.
The tuned model can then be used to
predict the production of charmonium states not considered in the
tuning step or in other kinematic regimes.

Production of the $\psi^\prime$  in proton induced collisions has been 
measured over a wide range of energies in several experiments at Fermilab
and CERN. But the experimental situation is still not satisfactory
since most measurements suffer either from limited sample size or from
additional large uncertainties resulting from signal extraction
in the presence of large backgrounds.
In many of the fixed-target experiments, 
absorber material was placed behind the target to 
reduce the muon background
produced by pion and kaon decays in flight.  
For such experiments the $e^+e^-$ mode was not observable 
and also the mass resolution suffered from the scattering of the muons
in the absorber.
In some cases, the mass resolution was insufficient for a 
separation of the $J/\psi$ and $\psi^\prime$ signals.

In contrast, using the
HERA-B detector's advanced particle identification and tracking systems,
it was possible to 
reduce the background from $\pi$ and $K$ decays in flight
by using track quality criteria
without inserting absorber material, and thereby to have
good mass resolution and also afford the possibility of triggering
on and reconstructing electrons.
The $\mu^+\mu^-$ and $e^+e^-$ decay channels were recorded
simultaneously. 
The rather large HERA-B samples of $J/\psi$ and $\psi^\prime$
(300,000 and 5000 events , respectively) and well
separated $J/\psi$ and $\psi'$ signals
will contribute to the clarification of the experimental situation.

In order to minimize possible systematic biases introduced by acceptance
and efficiency corrections, we report
the ratio of the $\psi'$ and $J/\psi$ production
cross section 
times the dilepton branching ratio.
The sample size allows measurements of
this ratio in several bins of the Feynman variable, $x_F$,
and transverse momentum, $p_T$.
The decay angular distribution of the ratio, which is related to 
the difference of the $\psi'$ and $J/\psi$ polarization
was also studied.

The branching ratio of $\psi'$ in the dimuon decay channel is
not well known.
The PDG value of the branching ratio $B'(\mu)=(0.73\pm0.08)\%$~\cite{pdg} 
for $\psi'$ has a relative error of $11\%$
and is a factor of 3 less accurate than the corresponding branching ratio
for the $e^+e^-$ decay, $B'(e)$. 
Furthermore almost all measurements of $\psi'$ to $J/\psi$ production ratio
have been performed in the $\mu^+\mu^-$ channel only.
The production ratio measured in $pp$ collisions by detecting
$e^+e^-$ pairs~\cite{isr} is rather imprecise $(0.019\pm0.007)$.
Our relatively large samples
in both channels therefore permit a sensitive test of
$e-\mu$ universality in $\psi'$ dilepton decays.  

The paper is structured as follows. After a short description of the 
HERA-B detector, the trigger and the data sample in Sect.\,\ref{detector}, 
the method used for the measurement of 
the $\psi'$ and $J/\psi$ production ratio
is introduced in Sect.\,\ref{method}.  
The Monte Carlo simulation and the event reconstruction and 
selection are presented 
in Sects.\,\ref{montec} and \ref{select}. 
In Sect.\,\ref{distr_cos}, the difference of $\psi'$ and $J/\psi$ 
polarization is studied, followed by a discussion of sources of 
systematic uncertainties in Sect.\,\ref{syst}. Finally, the experimental
results are presented, the differential $\psi'$ and 
$J/\psi$ production ratio 
in Sect.\,\ref{distr_xf}, 
the inclusive
$\psi'$ and $J/\psi$ production ratios in Sect.\,\ref{results}
and the nuclear dependence 
in Sect.\,\ref{distr_adepend}. 
Sect.\,\ref{compare} confronts the 
HERA-B results with previous measurements and theoretical model 
predictions. The paper ends with conclusions given in Sect.\,\ref{concl}.

\section{\boldmath Detector, trigger and data sample} 
\label{detector} 
The fixed target experiment HERA-B~\cite{lohse,hartoni}
was operated at the HERA storage ring at DESY until 2003.
The forward spectrometer of HERA-B covered angles 
from 15 to 220~mrad
in the bending plane  and from 15 to 160\,mrad vertically.
The target~\cite{ehret} consisted of 
two independent stations, separated longitudinally by 4 cm along the beam.
Each station contained four wires 
which could be independently positioned
in the halo of 
the HERA proton beam (inner, outer, below and above with respect 
to the beam).
Various materials (C, Ti, W) were used for the wires.  

The tracking system consisted of the Vertex Detector System (VDS) followed by
a magnetic dipole field of 2.13 Tm and a main tracker. 
The VDS~\cite{vds} was a silicon strip detector with 
a pitch of $50\,\mu{\rm m}$.
The Inner Tracker (ITR)~\cite{itr} of the main tracking system used
micro-strip gas chambers with 300\,$\mu$m pitch 
in the inner part of the main tracker up to distances of 25 cm
around the proton beam.
The Outer Tracker (OTR)~\cite{otr1,otr2} used honeycomb drift chambers 
with 5 and 10\,mm diameter cells covering the remaining acceptance.
A spatial resolution of about $150\,\mu$m and $360\,\mu$m was achieved for the ITR
and OTR, respectively.   

The main tracker system consisted of four stations
immediately behind the magnet:
the Pattern Chambers (PC), and 
two stations positioned further downstream:
the Trigger Chambers (TC).
A Ring Imaging Cherenkov 
detector (RICH)~\cite{rich}
was situated between the PC and TC stations.
The TC chambers were mainly used by the trigger. 
The TC also served together with the PC 
to provide additional measurements of the track position and 
direction. 
The number of layers was 30 and 12 in the PC and TC chambers, respectively. 

Particle~~~identification~~~was~~~provided~~by~~~the~~~RICH,
a MUON~~system and~~an Electromagnetic~~Calorimeter (ECAL).
The ECAL~\cite{ecal} followed by the MUON system were located at the
end of the spectrometer. 
The ECAL was optimized for good electron/gamma energy resolution
and electron--hadron discrimination. 
The ECAL had variable granularity:
the cell sizes were 2.2\,cm, 5.5\,cm and 11\,cm in the inner, middle and 
outer part, respectively.
The ECAL was instrumented with
a fast digital read-out and a pretrigger system.
The MUON system~\cite{muon} consisted of four tracking stations 
at various depths in an iron or concrete absorber.
Only muons above 5\,GeV/$c$ have a significant probability of penetrating
through the absorbers.

The trigger chain included the ECAL and MUON pretrigger systems 
which provided lepton candidate seeds
for the First Level Trigger (FLT) which in turn required at
least two pretrigger seeds.
Starting from the seeds, the FLT attempted to find tracks in a subset of
the OTR tracking layers, requiring that at least one of the seeds
results in a track. 
Starting once again from the pretrigger seeds, the
Second Level Trigger (SLT) 
searched for tracks using all OTR layers and continued the tracking through
the VDS, finally requiring at least two fully reconstructed tracks which
were consistent with a common vertex hypothesis.
The pretriggers and FLT are hardware
triggers which reduce with an output rate of typically 25 kHz
while the SLT is a software trigger running on a farm of 240 Linux PCs.

This analysis is based on 164 million dilepton triggered events collected  
in the 2002--2003 physics runs. About 300,000 $J/\psi$ lepton 
decays contained
in the data sample are almost equally divided between the dimuon and
dielectron channels. The wire materials used are
carbon ($A=12$, about $65\%$ of the $J/\psi$ data),
tungsten ($A=184$, about $31\%$) and titanium ($A=48$, about $4\%$).  
Only runs in which detector components performed well and
trigger conditions were stable were used.
Six periods with constant experimental
conditions are defined and the data are grouped accordingly.

\section{\boldmath Measurement method}  
\label{method}  
The analysis is based on the selection of dilepton events 
and fitting of the dilepton invariant mass
spectra in the area around the $J/\psi$ and $\psi'$ signals.
The number of events in the $J/\psi$ peak, $N_\psi$, is given by:
\[
  N_\psi\,=\,\sigma(J/\psi) \cdot B(J/\psi \to l^+l^-)
  \cdot {\cal L} \cdot \epsilon\,,
\]
i.e. the product of the cross section ($\sigma(J/\psi)$), 
the branching ratio into dilepton pairs ($B$), 
the integrated luminosity (${\cal L}$)
and the total reconstruction efficiency ($\epsilon$). 
The efficiency takes the following effects
into account: the probability for leptons to be within 
the detector acceptance and
to be properly registered in the detectors,
the probability for the dilepton pair to be triggered, 
the probabilities to reconstruct the tracks and the dilepton vertex, 
and the probability
to select the dileptons for the final analysis.
The luminosity is  identical for all charmonium states produced
on the same targets, and therefore cancels in the ratios.
The ratio of the $\psi'$ and $J/\psi$ cross sections in the $l^+l^-$ channel,
$R_{\psi'}(l)$, is equal to:   
\begin{equation}
R_{\psi'}(l)\,=\,
\frac{B'\cdot\sigma'}{B\cdot\sigma}\,=\,
\frac{N_{\psi'}}{N_{\psi}}
\cdot \frac{\epsilon}{\epsilon'}\,, 
\label{rpsi}
\end{equation}
where $l$ denotes the leptonic decay channel ($e$ or $\mu$),
$\sigma$($\sigma^\prime$)
is the $J/\psi$ ($\psi^\prime$) production cross section 
and $B$ ($B^\prime$)
is the branching ratio for the $l^+l^-$ decay 
of the $J/\psi$ ($\psi^\prime$) meson.

The ratio $N_{\psi'}\,/\,N_{\psi}$
or ``raw $\psi'/\psi$ ratio'',
is defined from the fit of the $J/\psi$ and $\psi'$ signals and 
must be corrected by the efficiency ratio, $\epsilon/\epsilon'$,
where $\epsilon$ is defined above and $\epsilon'$ 
is the corresponding efficiency for detection of $\psi'$ mesons.
The efficiencies are evaluated by  a Monte Carlo simulation.

For the analysis we select the kinematic domain 
\begin{equation}
-0.35 < x_F < 0.10\,\,\,
{\rm and}\,\,\,
0 < p_T < 4.5\,\,{\rm GeV}/c  
\label{kin_region}
\end{equation}
for the Feynman variable $x_F$ and the transverse momentum $p_T$,
respectively. 
This domain corresponds to our acceptance range in $x_F$ and at high $p_T$
is limited by lack of $\psi'$ events, i.e. is determined by the sample size.

\section{\boldmath Monte Carlo simulation}  
\label{montec} 
Results of Monte Carlo simulations (MC) were used to determine the 
efficiencies 
of $J/\psi$ and $\psi'$ triggering, reconstruction and selection.
$J/\psi$ and $\psi'$ production was simulated by generating the basic
process, $p{\rm N} \rightarrow c\bar{c}X$ 
with PYTHIA 5.7~\cite{pythia} and hadronizing the $c$ and $\bar{c}$ 
quarks with JETSET 7.4~\cite{pythia}.
The momentum and energy of the remaining hadronic system, $X$,
is given as input to FRITIOF 7.02~\cite{fritiof} which generates 
particles in the underlying event taking into account 
their interactions inside the nucleus.

The generated $\psi'$ events were assigned weights such that the resulting 
weighted $x_F$ distribution of the $\psi'$ matched that of the $J/\psi$,
since both measurements and theoretical 
models show little or no difference of these distributions 
in the kinematic domain Eq. (\ref{kin_region}) as 
described in Sect.~\ref{distr_xf}. 
Additional weights were applied 
to both $J/\psi$ and $\psi'$ events such that the resulting 
$p_T$ distribution matches
\begin{equation}
\frac{d\sigma}{dp_T^2} \propto 
{\left[ 1 +  
  {\left( \frac {35\,\pi\,p_T}
{256\,\langle p_T \rangle} \right)}^2 
\right]}^{-6},\,\langle p_T \rangle=1.29\,{\rm GeV}/c. 
\label{dndpt2} 
\end{equation} 
The average transverse momentum, $\langle p_T \rangle$, 
in (\ref{dndpt2}) was taken
from a preliminary HERA-B analysis of the $J/\psi$ sample~\cite{ptherab}.
The effects of possible differences in $p_T$ and $x_F$ distributions 
for the $J/\psi$ and $\psi'$ are described in Sect.\ref{syst}.

A GEANT 3.21~\cite{geant} based package~\cite{hbgean} 
performed tracking of the
particles through the HERA-B detector and the simulation of the detector 
response.    
The status (dead/alive), efficiency and noise level 
of each detector channel were inferred from the data separately for each
of six running periods (see Sect.~\ref{detector}) and used in modeling
the detector response.
The MC events were 
reconstructed with the same code used for the data.

\section{Event reconstruction and selection}   
\label{select}  
The track reconstruction included the following steps:
finding  straight track segments in the VDS and PC area,
propagation of the PC segments through the TC area,
matching of VDS and PC segments and a full iterative fit of the tracks.
The track segments were propagated into     
the TC area to provide a better measurement of the tracks   
in the RICH, ECAL and MUON detectors.
Particle identification estimators were evaluated for each track
using information from these three detectors. 

Pairs of like sign and unlike sign candidates of
muons and electrons were selected. 
A vertex fit was performed for
each candidate with a weak cut on the $\chi^2$ probability 
$(\ge\,\,10^{-5})$ to ensure that the tracks originated from the same vertex. 
At this stage, $99\%$ of the recorded $J/\psi$ and $\psi'$ candidates
with reconstructed lepton tracks were selected.
Further cuts and fits of the 
dilepton invariant mass spectra which were
specific to the muon and electron modes will be discussed
in the following two subsections.

\subsection{\boldmath Selection of $J/\psi,\,\psi' \to \mu^+\mu^-$}  
\label{select_mu}  
For the selection of $\mu^+\mu^-$ pairs, 
soft cuts were applied to
the momenta 
$(400\,{\rm GeV}/c > p > 6\,{\rm GeV}/c)$ and transverse momenta
$(p_T > 0.7\,{\rm GeV}/c)$ of the muons. The lower cut 
values were close
to the trigger requirements and the upper momentum cut  
rejected tracks outside the expected range for
$J/\psi(\psi')$ decay products. 

The main background in the muon channel was from muons produced in
$K,\pi \to \mu\nu$ decays in flight. If a decay occurred
inside the tracking system, the ``broken trajectory'' could
have led to a large $\chi^2$ in the track fit. A decay between
the VDS and the PC could additionally 
have led to a poor match
between VDS
and PC track segments.  
The tracks from the main tracker were projected
into the MUON system, where hits were associated with the track
and used to obtain 
a $\chi^2$--value with respect to the extrapolated track, taking into account
multiple scattering. 
This muon--$\chi^2$ was used to obtain
a quality estimator varying from 0 to 1 
and called the ``muon likelihood''.
Kaon and pion decays between the PC and the MUON system 
could have led to 
a large muon--$\chi^2$ and therefore a low muon likelihood,
as could wrongly assigning MUON hits 
produced by another 
particle.

\begin{figure}[htb]
\begin{center}
\epsfxsize=8.9cm
\epsfbox{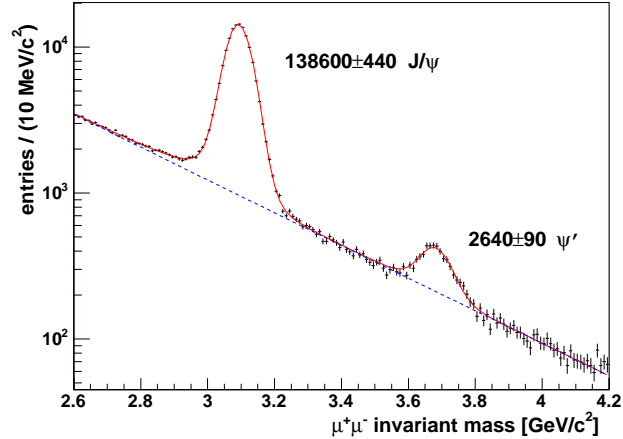}
\caption{\small
The invariant mass distribution of $\mu^+\mu^-$ 
in the region of the $J/\psi$ 
and $\psi'$ peaks for selected
dimuons from all three targets
and fitted with the sum of functions describing the signals   
and the background (dashed curve).}  
\label{goodsel}
\end{center}
\end{figure}
Thus, muon candidate tracks 
were selected according to the 
$\chi^2$ of their track fits,
qualities of their VDS--PC matches
and their muon likelihoods
in such a way that only about $5\%$ of real muons were discarded.
Tracks were also required to have a low probability of being a kaon
according to the RICH analysis.  

With these selections,
the background under the $J/\psi$ and $\psi'$ peaks
was reduced by a factor of 2.5.
The signal loss due to these selections  
was distributed rather evenly over the kinematic
range defined by Eq. (\ref{kin_region}) and was reproduced by the MC.
After the selection, $(89.4\pm0.2)\%$ of the events 
in the $J/\psi$ peak survived in the data and $(89.1\pm0.1)\%$ in the MC.   

The dimuon mass spectra for both
$J/\psi$ and $\psi'$ were described as
a sum of two functions~\cite{rdecay}: 1) a symmetric function,
being a superposition of three Gaussians, which takes into account
track resolution and
effects of Moli\`{e}re scattering,  
2) a function representing a radiative tail due
to the emission of a photon in the final state of the dimuon decay
$J/\psi(\psi') \to \mu^+\mu^-\gamma$.  

The presence of the radiative tails 
and tails from Moli\`{e}re scattering
in the invariant mass distribution of the $J/\psi$ and $\psi'$
signals demanded 
that both peaks be fit simultaneously.
The total number of events in a peak was obtained by an integration
of the fitted function over the full range of possible dimuon masses. 
The mass of the $\psi'$ was obtained by scaling
the fitted $J/\psi$ mass by the factor
${M_0(\psi^{\prime}) / M_0(J/\psi)}$, where the $M_0$ were the nominal
values of the respective masses. 
The fitted $J/\psi$ mass resolution 
was rescaled for the $\psi'$ 
using the dependence of the momentum resolution for the HERA-B spectrometer.
The background was described as an exponent
of a quadratic polynomial of the dimuon mass.

The functions describe the dimuon mass spectra well as shown 
in Fig.\ref{goodsel}, 
where $\chi^2=154$ for 153 degrees of freedom.
The fitted mass of $J/\psi$ is 3093 ${\rm MeV}/c^2$ and the 
mass resolution (FWHM\,/\,2.35) is 38 ${\rm MeV}/c^2$.   

\subsection{\boldmath Selection of $J/\psi,\,\psi' \to e^+e^-$ }  
\label{select_e}  
In the dielectron channel, the background was larger                         
than for muons mainly due to   
misidentification of charged pions interacting in the ECAL and
overlaps between energy deposits from photons and charged 
hadrons.    
Despite the higher background a careful study of the electron
identification cuts resulted in $J/\psi$ and $\psi^\prime$
signals of comparable purity and significance. 

The first step of the event selection consisted of
the requirement that  
track momenta lie between 4 and 400 GeV/$c$,
corresponding to the expected range for $J/\psi$ and $\psi^\prime$ decay
products.
Finally, an additional cut on the transverse energy of the ECAL cluster
($E_T> 1.15$ GeV) was applied in order to equalize different cut
thresholds used at the pretrigger level during the acquisition
periods. The signal selection was improved by rejecting  
electron/positron pairs having a distance of 
closest approach between the two tracks greater than 320 $\mu {\rm m}$.

The reconstructed momentum vectors of electrons and positrons 
were corrected for energy lost due to brems\-strahlung (BR) emission
in the material before the magnet (evaluated as about $18\%$ on average). 
BR was identified and determined 
for each track by looking for an energy deposition in coincidence with the
extrapolation of the track's VDS segment to the ECAL. The recovered energy was 
then added to the momentum measured by the tracking system. Being a clear 
signature of electrons, BR emission was also exploited    
to obtain a substantial background reduction, essential for an                  
accurate counting of the small $\psi^\prime$ signal. 
It was therefore required that at least one lepton of the decaying 
$J/\psi$ or $\psi^\prime$ had an associated BR cluster (defined as
an ECAL cluster close enough to the VDS segment extrapolation).
\begin{figure}[htb]
\begin{center}
\epsfxsize=8.9cm
\epsfbox{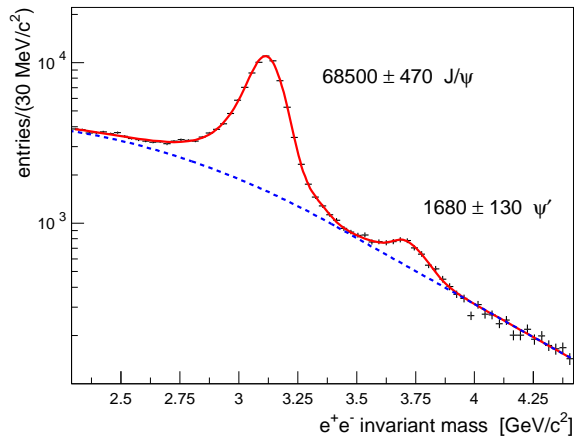}
\caption{\small
The invariant mass distribution of $e^+e^-$ in the region of the $J/\psi$
and $\psi'$ peaks for selected dielectrons from all three materials, 
fitted with the sum of the functions describing the signals and the 
background (dashed curve).}
\label{goodsel_ee}
\end{center}
\end{figure}

Two additional types of selection cuts were applied to those 
electrons and positrons which had no associated BR cluster.
The first of these cuts is on the $E/p$ ratio,
where $E$ is the energy of the electron/positron 
cluster and $p$ is the associated 
track momentum. The $E/p$ distribution for electrons has a Gaussian 
shape with mean value close to 1 and width, 
$\sigma_{E/p}$, varying between 6.4$\%$ and 7.4$\%$ depending on calorimeter 
section. Values of $E/p$ far from 1 correspond to particles, mostly
hadrons, which release only part of their energy in the calorimeter.
The second cut was applied 
to the distance between the reconstructed cluster and the
track position extrapolated to the ECAL ($\Delta_x$ and $\Delta_y$   
for the x and y direction). The $\Delta_x$ and $\Delta_y$
distributions for electrons are, apart from a small tail, well 
described by Gaussians centered at zero, with widths between 0.2\,cm 
and 1.0\,cm, depending on calorimeter section.
A cut on these quantities leads to a significant reduction of the 
contamination from hadrons and          
random cluster-track matches, for which the $\Delta_{x(y)}$
distributions are expected to be significantly wider compared 
to those of the electrons. 

All the above requirements were simultaneously optimized by
maximizing the significance of the $\psi^\prime $ MC signal. 
 The accepted ranges for $E/p$, $\Delta_x$ and $\Delta_y$ were respectively 
 determined as  $-1.0\cdot\sigma_{E/p} < E/p - 1 < 3.2\cdot\sigma_{E/p}$,
$|\Delta x | < 1.7\, \sigma_x$ and $|\Delta y | < 1.8\, \sigma_y$.
As a result,                  
the $S/B$ ratio, evaluated for the  $J/\psi$ (the only one of the two 
charmonium states visible also before the selection cuts), 
increased by about a factor of $10$ with respect to the triggered events,
with an efficiency for signal selection of 
$(34 \pm 3)\%$ in the real data and $(35.1 \pm 0.4)\%$ in the MC simulation.

To count the number of $J/\psi$ and $\psi^\prime$,
a Gaussian for the right part of the peak
and a Breit-Wigner for the left part were used, allowing for
a sizeable asymmetry of the signal due to the not fully
reconstructed BR emission and a contribution from $J/\psi \to e^+e^- \gamma$
decay. 
Both the width and the asymmetry of the
$\psi^\prime$ signal were kept at a fixed ratio (determined from the MC)
with respect to those of the  $J/\psi$.     
The background was parametrized as a Gaussian for the                    
low invariant mass values and an exponential at higher mass, 
requiring continuity of the resulting function and its first derivative.
The invariant mass distribution plotted in Fig.~\ref{goodsel_ee}
shows our final selection of $68500 \pm 470$  $J/\psi$  and  $1680 \pm 130$
$\psi^\prime$: the resulting mass and width (of the Gaussian) 
of the $J/\psi$ are, respectively, $3110$ and $72$~MeV/$c^2$.
The $\chi^2$ of the fit is 89 for 69 degrees of freedom.

\section{\boldmath Angular dependence of $R_{\psi'}$}      
\label{distr_cos}      
Measurements of polarization (the commonly used term to denote spin alignment)
provide one of the most significant tests
of models of charmonia production~\cite{benhadr}.
Understanding polarization at moderate and low $p_T$ 
is also important since most of
the published $p_T$-integrated data come from the lower $p_T$ region.  
The polarization of charmonium 
is measured by observation of the angular 
distribution in its decay into $l^+l^-$: 
\begin{equation}
\frac{d\sigma}{d{\rm cos}\theta} \,\propto\, 
(1~+~\lambda\,\rm{cos}^2\,\theta),
\label{dndtheta} 
\end{equation} 
where $\theta$ is the polar angle of the $l^+$ with respect to the $z$ axis of
a polarization frame defined in the rest system of the charmonium state.
In general, the parameter $\lambda$ depends on the definition of
the polarization frame~\cite{benhadr}. 
The commonly used polarization frames
are specified by the choice of the $z$ axis.
The three-momentum of the beam particle, $\vec{p_b}$, 
and that of the target, $\vec{p_t}$, in the rest system of the charmonium
can be used to define the $z$ axis.

In the Gottfried-Jackson frame, the $z$ axis is parallel to $\vec{p_b}$,
in the $s$-channel helicity (or recoil) frame,
the $z$ axis is defined as the direction of the charmonium three-momentum
in the hadronic center-of-mass frame, i.e. it is parallel to
$-(\vec{p_b}+\vec{p_t})$, and 
in the Collins-Soper frame, the $z$ axis bisects the angle
between $\vec{p_b}$ and $-\vec{p_t}$.

For $J/\psi$ production, several measurements of polarization
have been published.
The parameter $\lambda$ measured in $p$Be collisions~\cite{e672p} 
in the Gottfried-Jackson frame is
$\lambda=0.01\pm0.12\pm0.09$ and $\lambda=-0.11\pm0.12\pm0.09$
at 530 GeV and 800 GeV, respectively.
A high statistics measurement~\cite{e866pol,e866phd} 
in 800 GeV $p$Cu interactions  
(Collins-Soper frame) resulted in $\lambda=0.069\pm0.004\pm0.08$.
The quoted errors are statistical and systematic.
A measurement~\cite{heinrich} for $\psi'$
was performed for $\pi^-{\rm W}$ interactions at 253 GeV
where $\lambda'=0.02\pm0.14$  was obtained
in the Gottfried-Jackson frame. 
All results are consistent with 
a small difference in the $\psi'$ and $J/\psi$ polarization.
The collider result~\cite{cdf} ($s$-channel helicity frame)
$\lambda'=-0.08\pm0.63\pm0.02$
for a mean $p_T$ equal to 6.2 GeV/$c$ is also consistent with 
that,
but statistically less significant
and, because of the relatively high $p_T$, 
not directly comparable with the fixed-target results.

In the kinematic domain (\ref{kin_region}) of HERA-B,
the $\psi'$ signal is clearly visible 
in the limited interval $|\cos\,\theta| < 0.6(0.8)$
for electrons (muons) 
in the Gottfried-Jackson frame.
In the $s$-channel helicity frame the acceptance is more uniform 
as a function of $\cos\,\theta$
and $R_{\psi'}$ is measurable over
the full interval of $\cos\,\theta$. 
We therefore prefer to present our results in the $s$-channel helicity frame.
The values of $R_{\psi'}$ for ten $\cos\theta$ bins are
shown in Table~1.
\begin{figure}[h]
\begin{center}  
\epsfxsize=8.9cm 
\epsfbox{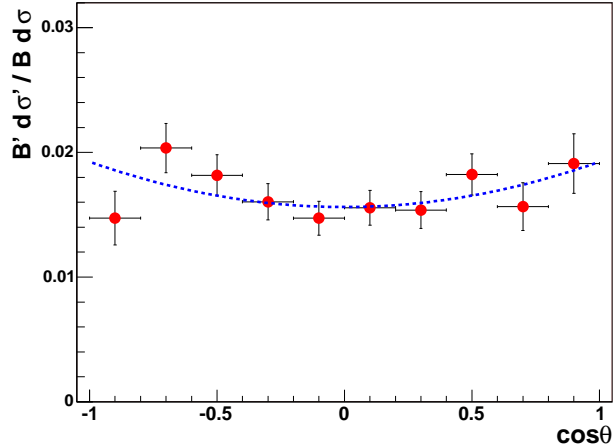} 
\caption{\small 
$R_{\psi'}$ as a function of $\cos\theta$ 
in the $s$-channel helicity frame. 
The dashed curve is the fitted function (\ref{drdtheta})
for $\lambda = 0$ and the fitted parameter $\lambda'=0.23\pm0.17$.}  
\label{rall_vs_cs} 
\end{center} 
\end{figure} 
\begin{table}[h] 
\begin{center} 
\caption{The ratio of $\psi'$ and $J/\psi$ cross sections 
in the muon, $R_{\psi'}(\mu)$, electron, $R_{\psi'}(e)$, channels 
and for both channels combined, $R_{\psi'}$,  
obtained for different $\cos\,\theta$ intervals 
using the data from all targets. 
The value $\theta$ is the polar angle of the positive lepton 
in the $s$-channel helicity frame.
Additional systematic uncertainties of    
$5\%$ for the electron channel and $3\%$ for 
both the muon channel and for the combined result 
(see Table~\ref{systsum}) are not included here.}
\begin{tabular}{lccc} 
\hline\noalign{\smallskip} 
$~~~~~{\rm cos}\,\theta$ & $R_{\psi'}(\mu)$ & $R_{\psi'}(e)$ & $R_{\psi'}$ \\ 
\noalign{\smallskip}\hline\noalign{\smallskip} 
$-1.0\div-0.8 $& $0.0148\pm0.0024$ & $0.0144\pm0.0054$&$0.0147\pm0.0022$\\ 
$-0.8\div-0.6 $& $0.0211\pm0.0022 $ &$0.0169\pm0.0046$&$0.0203\pm0.0020$\\ 
$-0.6\div-0.4 $& $0.0203\pm0.0019 $ &$0.0107\pm0.0034$&$0.0180\pm0.0017$\\ 
$-0.4\div-0.2 $& $0.0154\pm0.0016 $ &$0.0185\pm0.0033$&$0.0160\pm0.0014$\\ 
$-0.2\div~~0.0 $ & $0.0152\pm0.0016$ &$0.0130\pm0.0028$&$0.0147\pm0.0014$\\ 
$~~0.0\div~~0.2 $  & $0.0161\pm0.0016 $ &$0.0140\pm0.0028$&$0.0156\pm0.0014$\\ 
$~~0.2\div~~0.4 $  & $0.0152\pm0.0017 $ &$0.0159\pm0.0030$&$0.0154\pm0.0015$\\ 
$~~0.4\div~~0.6 $  & $0.0187\pm0.0019 $ &$0.0167\pm0.0035$&$0.0182\pm0.0017$\\ 
$~~0.6\div~~0.8 $  & $0.0157\pm0.0021 $ &$0.0154\pm0.0043$&$0.0156\pm0.0019$\\ 
$~~0.8\div~~1.0 $  & $0.0176\pm0.0026 $ &$0.0269\pm0.0060$&$0.0191\pm0.0024$\\ 
\noalign{\smallskip}\hline 
\end{tabular} 
\end{center} 
\label{table_cos}       
\end{table} 

The dependence of $R_{\psi'}$ on $\cos\,\theta$
can be described as 
\begin{equation}
\frac{B'\,\,d\sigma'/d\cos\theta}
{B\,\,d\sigma/d\cos\theta}\,\propto\,
\frac{1~+~\lambda'\,\rm{cos}^2\,\theta}{1~+~\lambda\,\rm{cos}^2\,\theta},
\label{drdtheta}
\end{equation}
where $\lambda'$ and $\lambda$ are related to the polarization 
of the $\psi'$ and the $J/\psi$, respectively.
The fit of $R_{\psi'}$  (see Fig. \ref{rall_vs_cs})
with this function, where $\lambda'$ is a free parameter  
and $\lambda$ fixed, depends on the $\lambda$ value, but
the difference $\lambda'-\lambda$ is nearly constant.
We obtain in the $s$-channel helicity frame
\begin{equation} 
  \Delta \lambda = \lambda' - \lambda = 0.23\pm0.17,  
\label{dlambda}  
\end{equation}  
for the fit with $\lambda=0$.
To check the stability of the result, we varied
$\lambda$ in the interval ($-0.2$, $0.2$), 
which is broader than the limits given  
by the published measurements of $J/\psi$ polarization,
and obtained a variation of $\Delta \lambda$
in the range $\pm0.04$, which is
much smaller than the statistical uncertainty.

\section{\boldmath Systematic errors of $R_{\psi'}$}     
\label{syst}     
The correction of the raw $\psi'/\psi$ ratio in (\ref{rpsi}) using 
the efficiency ratio, $\epsilon/\epsilon'$, calculated 
by MC was the subject of a detailed study.
The efficiency ratio, $\epsilon/\epsilon'$, was mainly determined 
by geometric factors since effects such as detector 
and trigger efficiencies nearly cancel out in the ratio.
The geometrical factors can be calculated reliably with 
MC and were found to be quite stable. 
This is confirmed by the data since the raw $\psi'/\psi$ ratio
is also quite stable for different periods of
data taking and target configurations.

Nevertheless, a small systematic bias can appear due to  
uncertainties in the simulation of kinematical distributions
of charmonia production as well as of the detector and trigger.
Possible biases were evaluated by varying parameters of the MC simulation.  

The MC simulation assumed unpolarized production 
for $J/\psi$ and
$\psi^\prime$.  However, due to
the limited acceptance, the
efficiency ratio $\epsilon/\epsilon'$ in Eq.~(\ref{rpsi})
could depend on a difference between the simulated decay angular
distributions for the $\psi^\prime$ and $J/\psi$.
The ratio $\epsilon/\epsilon'$ 
depends slightly, almost linearly on 
the difference $\lambda' - \lambda$ and consequently
\begin{equation} 
\frac{\Delta R_{\psi'}}{R_{\psi'}} = 0.2\,\cdot\,\Delta\lambda.
\label{drrvsdl} 
\end{equation} 
To obtain Eq.~\ref{drrvsdl}, we evaluated the efficiency ratio
$\epsilon/\epsilon'$ by simulating $\psi'$ and $J/\psi$ production 
for different $\Delta\lambda$ values 
in the Gottfried-Jackson frame.
In this frame, the acceptance is less uniform as a function 
of $\cos\theta$, and the bias of $R_{\psi'}$ due to the
possible difference of angular distributions for 
the $\psi^\prime$ and $J/\psi$ 
is dominated by the size of $\Delta\lambda$.
Theoretical arguments suggest that the  
difference in the 
$\psi'$ and $J/\psi$ polarization is small.
The color evaporation model (CEM) predicts 
that $\lambda$ and $\lambda'$ are equal~\cite{yelrep}. 
In the Nonrelativistic QCD (NRQCD) approach, the polarization of the $J/\psi$ 
is larger than that of the $\psi'$ due to feeddown from the $\chi_c$,
but $\lambda$ is unlikely to exceed $\lambda'$ by more than
0.2~\cite{benhadr}.
For the HERA-B experiment, 
the values for $\lambda$
calculated in the NRQCD approach~\cite{yelrep} range from 0 to 0.1.
We estimated the contribution to the relative systematic error due to 
polarization $\sigma_{sys}(R_{\psi'})/R_{\psi'}\,=\,0.02$, corresponding
to $\Delta\lambda = 0.1$ in (\ref{drrvsdl}).

Theoretical predictions for $J/\psi$ and $\psi'$ polarization
in fixed-target experiments
are largely independent of the target and beam types 
and vary slowly with beam
energy~\cite{yelrep}. Therefore 
direct comparison of results of different experiments 
is meaningful.
Combining measurements
\cite{e672p,e866pol,e771pol} for $\lambda$,
 \cite{heinrich} for $\lambda'$ and the HERA-B result~(\ref{dlambda}), 
we obtained $\Delta\lambda = 0.09\pm0.11$.
The latter estimate disfavors also a big difference
between $\lambda$ and $\lambda'$. 
\begin{table*}[ht] 
\begin{center} 
\caption{
Estimated relative systematic errors of the $R_{\psi'}$ 
in the $\mu^+\mu^-$, $R_{\psi'}(\mu)$, and $e^+e^-$, $R_{\psi'}(e)$, channels
from various sources, as listed.
The total uncertainty was obtained by quadratic summation.}
\label{systsum} 
\begin{tabular}{ccc} 
\hline\noalign{\smallskip}  
 Source of uncertainty 
& $\sigma_{\rm sys}(R_{\psi'}) / R_{\psi'}(\mu)$ &
$\sigma_{\rm sys}(R_{\psi'}) / R_{\psi'}(e)$ \\ 
\hline\noalign{\smallskip}  
Polarization        & 0.02 & 0.02 \\
$p_T$ distribution  & 0.0006 & 0.0006\\  
$x_F$ distribution  & 0.004 & 0.004\\ 
Trigger simulation  & 0.006 & 0.02\\
Counting method    & 0.014 & 0.035 \\
\hline\noalign{\smallskip}
Total               & 0.03 & 0.05 \\
\hline\noalign{\smallskip} 
\end{tabular} 
\end{center} 
\end{table*} 

To estimate the importance of the MC model used for the production
kinematics, the parameters of the generated $x_F$ and $p_T$ distributions
were changed within variations allowed by the measured differences between
$\psi^\prime$ and $J/\psi$ 
kinematics~\cite{ptherab,e771,e705}. 
The corresponding variations of $R_{\psi'}$
were found to be rather small as listed in Table \ref{systsum}.
Contributions to the systematic error due to differences 
of kinematical distributions and polarization of $\psi'$ 
and $J/\psi$ were highly correlated for muons and electrons. 

Uncertainties connected to the trigger
were evaluated by removing the trigger simulation for both muons and
electrons and, as a second test, increasing the value  
of the $E_T$ threshold with respect to the 
one used by the pretrigger for electrons. 
These are quite extreme tests which surely overestimate possible trigger
biases, nonetheless, 
as seen in Table~\ref{systsum},
the changes in the $R_{\psi'}$ do not exceed
$1\%$ and $2\%$,  for muons and electrons,
respectively.
Moreover, it has been verified for electrons that any systematic 
bias due to
the BR request on the value of $R_{\psi^\prime}$ is negligible.

The last item listed in Table~\ref{systsum} 
is related to different methods of 
counting the numbers of $J/\psi$ and $\psi^\prime$ 
in~(\ref{rpsi}). 
We obtained these values either by taking all events in the kinematic
domain~(\ref{kin_region}) or by summing the events in the bins
of the $x_F$, $p_T$ or $\cos \theta$ distributions. 
In doing so,
the efficiencies were evaluated either for the whole kinematic 
domain or for the different bins.    

The total relative systematic error 
on $R_{\psi'}$ was estimated to be 
$3\%$ and $5\%$ for the muon and electron modes, respectively.
The larger error for the electron mode can be attributed to
the higher contribution of background.

\section{\boldmath $x_F$ and $p_T$ dependence of $R_{\psi'}$}    
\label{distr_xf}    
To obtain the ratio of the $\psi'$ to $J/\psi$ production cross sections
times the dilepton branching ratio, $R_{\psi'}$,
as a function of $x_F$ we selected dilepton pairs in eight $x_F$ 
intervals and performed the full analysis for each interval
independently. 
The results are listed in Table~3, separately for $\mu^+\mu^-$, $e^+e^-$
channels as well as for the combined result.
\begin{figure}[h]
\begin{center}
\epsfxsize=0.70\textwidth
\epsfxsize=8.9cm
\epsfbox{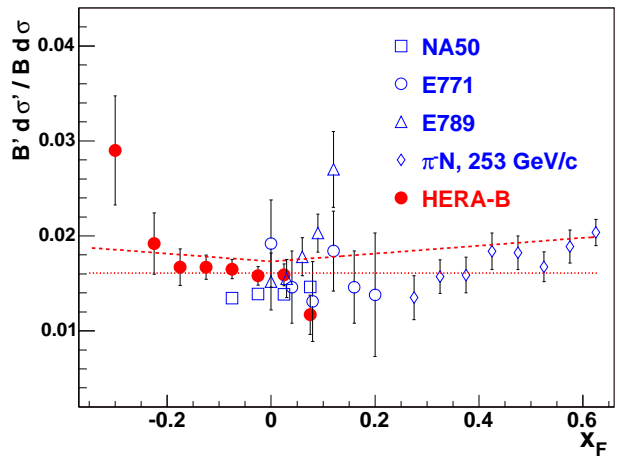}
\caption{\small
Measurements of $R_{\psi'}$
as a function of  $x_F$. 
Results of this work, combined for $\mu$ and $e$ decay modes and
for all targets, are shown together
with previous experiments which measured charmonia production
in $p$N (NA50~\cite{na50}, E771~\cite{e771}, E789~\cite{e789}) and
$\pi^-$N interactions~\cite{heinrich}.
Only statistical errors are shown.
Two calculations, one 
using the NRQCD approach~\cite{rvogt} 
and the other using
the CEM model~\cite{rvogt2}, are
indicated by the dashed and dotted curves, respectively.
Both calculations were made for $pp$ interactions and corrected
for the nuclear suppression in the HERA-B target  
as described in the text.}
\label{rall_vs_xf}
\end{center}
\end{figure}
\begin{table*}
\begin{center}
\caption{The ratio of $\psi'$ and $J/\psi$ cross sections
in the muon, $R_{\psi'}(\mu)$, electron, $R_{\psi'}(e)$, channels
and for both channels combined, $R_{\psi'}$, 
obtained for intervals in $x_F$  using the data from all targets.
Additional systematic uncertainties of
$5\%$ for the electron channel and $3\%$ for both
the muon channel and for the combined result (see Table~\ref{systsum})
are not included here.}
\begin{tabular}{lccc}
\hline\noalign{\smallskip}
~~~~~~~$x_F$ &  $R_{\psi'}(\mu)$ & $R_{\psi'}(e)$ & $R_{\psi'}$ \\
\noalign{\smallskip}\hline\noalign{\smallskip}
$-0.35\div-0.25$ & $0.0309\pm0.0076$ & $0.0265\pm0.0088 $&$0.0290\pm0.0058$\\
$-0.25\div-0.20$ & $0.0172\pm0.0038$ & $0.0244\pm0.0061$ &$0.0192\pm0.0032$\\
$-0.20\div-0.15$ & $0.0156\pm0.0022$ & $0.0202\pm0.0040$ &$0.0167\pm0.0019$\\
$-0.15\div-0.10$ & $0.0165\pm0.0014$ & $0.0175\pm0.0030$ &$0.0167\pm0.0013$\\
$-0.10\div-0.05$ & $0.0162\pm0.0011$ & $0.0179\pm0.0025$ &$0.0165\pm0.0010$\\
$-0.05\div~~0.00$  & $0.0160\pm0.0011$ &$0.0152\pm0.0022$ & $0.0158\pm0.0010$\\
~~$0.00\div~~0.05 $  & $0.0162\pm0.0013$ &$0.0147\pm0.0025$&$0.0159\pm0.0012$\\
~~$0.05\div~~0.10$   & $0.0095\pm0.0025$ &$0.0165\pm0.0037$&$0.0117\pm0.0021$\\
\noalign{\smallskip}\hline
\end{tabular}
\end{center}
\label{table_xf}       
\end{table*}
Fig.~\ref{rall_vs_xf} displays the HERA-B results together with 
earlier measurements.
Neither the HERA-B nor the NA50 data~\cite{na50}
exhibit significant $x_F$ dependencies.
$R_{\psi'}$ measurements of E771~\cite{e771}    
in $p$N interactions  
extend up to $x_F=0.2$, but suffer from 
large statistical errors.
A statistically more accurate measurement, but
performed in  $\pi^-$N interactions~\cite{heinrich},
also indicates only a slow variation of $R_{\psi'}$ in
the range of low and moderate $x_F$ values. 
We may expect a
similarity of the $x_F$ behavior of the $R_{\psi'}$ in 
$p$N and $\pi^-$N interactions,
because of the dominance of gluon fusion in the charmonium production
process.

In the CEM~\cite{fritzsch} the differential
and integrated charmonium production rates for different charmonium states
should be proportional
to each other and independent of projectile, target and energy. 
The results of the CEM for $pp$ interactions~\cite{rvogt2} 
show a flat $x_F$ behavior and  
agree with our results (Fig.~\ref{rall_vs_xf}).
The NRQCD calculation~\cite{rvogt} of $\psi'$ and $J/\psi$ production in $pp$
interactions at 920 GeV shows a slow variation
of $R_{\psi'}$ versus $x_F$ and is consistent with our measurements as well.
The results of the NRQCD and CEM models were both corrected 
for nuclear suppression as will be
discussed in Sect.~\ref{compare}.

\begin{figure}[h] 
\begin{center}   
\epsfxsize=8.9cm  
\epsfbox{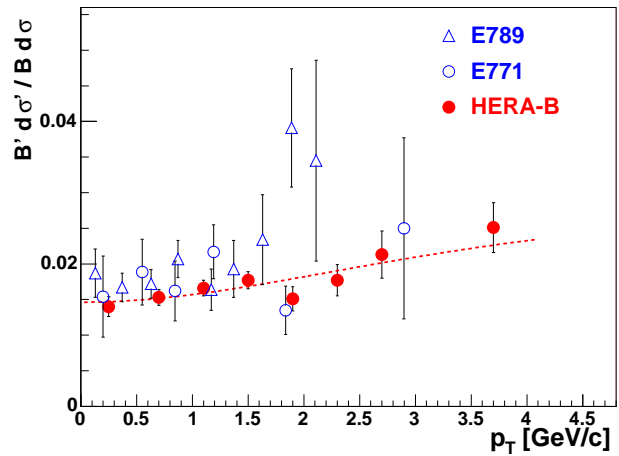}  
\caption{\small  
Measurements of $R_{\psi'}$
as a function of    
the transverse momentum.  
Combined results for $e$ and $\mu$ modes obtained for  
all three targets are shown together  
with previous results from experiments E771~\cite{e771} 
and E789~\cite{e789}. Only statistical errors are shown.
A fit of the HERA-B results 
with the ratio of functions (\ref{dndpt2})
as described in the text,
is shown by the dashed line.}
\label{rall_vs_pt}  
\end{center}  
\end{figure}

The results for $R_{\psi'}$ for eight bins in transverse momentum
are given in Table~4.
$R_{\psi'}$ shows a tendency to increase as
a function of $p_T$ as seen in Fig.~\ref{rall_vs_pt}.
The fit of $R_{\psi'}$ 
with the ratio of functions of the form given in
Eq. (\ref{dndpt2}),
where the average transverse momentum,  ${\langle p_T \rangle}'$, for $\psi'$
is a free parameter and ${\langle p_T \rangle}$ for $J/\psi$
is fixed to the value (\ref{dndpt2}), gives
${\langle p_T \rangle}'-{\langle p_T \rangle}=(0.08\pm0.03)\,{\rm GeV}/c$.
These results are consistent with those 
of E771~\cite{e771} and E789~\cite{e789},
although the large statistical errors of these two previous 
measurements do not allow for a decisive test.

\begin{table*}[h] 
\begin{center} 
\caption{The ratio of $\psi'$ and $J/\psi$ cross sections 
in the muon, $R_{\psi'}(\mu)$, electron, $R_{\psi'}(e)$, channels 
and for both channels combined, $R_{\psi'}$, 
obtained for intervals in $p_T$ using the data from all targets.
Additional systematic uncertainties of 
$5\%$ for the electron channel 
and $3\%$ for both the muon channel and for the combined result 
(see Table~\ref{systsum}) are not included here.} 
\begin{tabular}{lccc} 
\hline\noalign{\smallskip} 
$p_T$\,\,[GeV/$c$] &  $R_{\psi'}(\mu)$ & $R_{\psi'}(e)$ & $R_{\psi'}$ \\ 
\noalign{\smallskip}\hline\noalign{\smallskip} 
$0.0\div0.5$ & $0.0143\pm0.0015$ &$0.0129\pm0.0031$&$0.0140\pm0.0014$\\ 
$0.5\div0.9$ & $0.0156\pm0.0012$ &$0.0145\pm0.0022$&$0.0153\pm0.0011$\\ 
$0.9\div1.3$ & $ 0.0167\pm0.0012$&$0.0159\pm0.0025$&$0.0166\pm0.0011$\\ 
$1.3\div1.7$  & $0.0171\pm0.0014$&$0.0201\pm0.0027$&$0.0177\pm0.0012$\\ 
$1.7\div2.1$  & $0.0156\pm0.0019$&$0.0134\pm0.0034$&$0.0151\pm0.0017$\\ 
$2.1\div2.5$  & $0.0192\pm0.0027$&$0.0145\pm0.0040$&$0.0177\pm0.0022$\\ 
$2.5\div2.9$  & $0.0248\pm0.0041$&$0.0150\pm0.0055$&$0.0213\pm0.0033$\\ 
$2.9\div4.5$  & $0.0277\pm0.0043$&$0.0198\pm0.0061$&$0.0251\pm0.0035$\\ 
\noalign{\smallskip}\hline 
\end{tabular} 
\end{center} 
\label{table_pt}       
\end{table*} 

\section{\boldmath Combined results for $R_{\psi'}$ and for the 
double ratio of leptonic branching fractions}
\label{results}      
An independent analysis was performed for 
each target sample (C, Ti, W) and separately for 
both dimuon and dielectron
trigger modes. For each sample 
the dilepton invariant mass
spectra were fitted to obtain the raw $\psi'/\psi$ ratio
which was then corrected according to (\ref{rpsi}) by
the efficiency ratio $\epsilon/\epsilon'$.

The values of $R_{\psi'}$ 
in both dilepton channels  
obtained for all three targets are shown in Table~5.
For each target  
the values of $R_{\psi'}$ for the muon and electron mode were combined 
and the ratio $R_{\psi'}(\mu)/R_{\psi'}(e)$ was evaluated.
\begin{table*}[h]
\begin{center}
\caption{The ratio of $\psi'$ and $J/\psi$ production cross sections
per nucleus in the muon, $R_{\psi'}(\mu)$, electron, $R_{\psi'}(e)$, channels
and for both channels
combined, $R_{\psi'}$, for each target.
The ratio $R_{\psi'}(\mu) / R_{\psi'}(e)$ is also presented.
The quoted errors indicate the statistical and systematic errors,
respectively.}
\begin{tabular}{lcccc}
\hline\noalign{\smallskip}
 Target &  $R_{\psi'}(\mu)$ & $R_{\psi'}(e)$ & $R_{\psi'}$ &
$R_{\psi'}(\mu) / R_{\psi'}(e)$\\
\noalign{\smallskip}\hline\noalign{\smallskip}
C & $0.0166\pm0.0007\pm0.0004$&$0.0154\pm0.0012\pm0.0007$ &
$0.0163\pm0.0006\pm0.0005$ &  $1.08\pm 0.10 \pm 0.04$ \\

Ti & $0.0198\pm0.0029\pm0.0005$ & $0.0200\pm0.0045\pm0.0009$ &
$0.0199\pm0.0024\pm0.0006$ & $0.99\pm 0.27 \pm 0.04$\\

W & $0.0158\pm0.0011\pm0.0004$ & $0.0182\pm0.0025\pm0.0012$ &
$0.0162\pm0.0010\pm0.0005$ &  $0.87\pm 0.13 \pm 0.03$\\

\noalign{\smallskip}\hline
\end{tabular}
\end{center}
\label{alltarget}       
\end{table*}

\begin{figure}[htb]
\begin{center}
\epsfxsize=8.9cm
\epsfbox{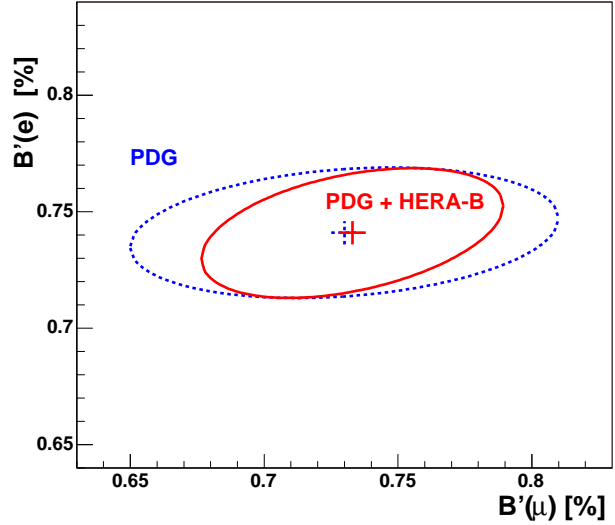}
\caption{\small
The error ellipse for $B'(e)$ versus $B'(\mu)$,  
evaluated by using the covariance 
matrix from~\cite{pdg} (dashed curve) 
and after (solid curve) applying
the HERA-B constraint given by (\ref{rpsiherab}).
}
\label{contours}
\end{center}
\end{figure}
We obtained the ratio
$R_{\psi'}(\mu)/R_{\psi'}(e)$, combined for all targets,
as a weighted average of the three corresponding values
from  Table~5:
\begin{equation} 
R_{\psi'}(\mu)/R_{\psi'}(e) = 1.00\pm0.08\pm0.04.
\label{rpsiherab}  
\end{equation}
The quoted errors are statistical and systematic.  
The systematic error was estimated by comparing the results of
different counting methods 
and by taking into account the uncertainty of trigger 
simulation (mainly due to the electron mode)
as discussed in Sect.~\ref{syst}.
Contributions to the systematic error due to differences
of kinematical distributions and polarization of $\psi'$
and $J/\psi$ almost cancel out in (\ref{rpsiherab}).

The ratio (\ref{rpsiherab}) also represents a direct 
measurement for the double 
ratio of the branching fractions
of the $\psi'$ and $J/\psi$ dilepton decays:
\begin{equation}
\frac {R_{\psi'}(\mu)}{R_{\psi'}(e)}\,=\,
\frac{B'(\mu) \cdot B(e)} {B(\mu) \cdot B'(e)}.
\label{rpsime}
\end{equation}
The PDG values of branching ratios are~\cite{pdg}:
$B(\mu)=(5.88\pm0.10)\%$, $B(e)=(5.93\pm0.10)\%$ for $J/\psi$ dilepton
decays and 
$B'(\mu)=(0.73\pm0.08)\%$, $B'(e)=(0.741\pm0.028)\%$ for 
$\psi'$ decays.
The $\psi'$ branching ratios are not independent since they are 
obtained by   
a global fit performed by the PDG~\cite{pdg},
which results in $B'(\mu)/B'(e)=0.99\pm0.13$.
The latter value and the $J/\psi$ branching ratios
correspond to the following determination of the double ratio:
\begin{equation}
\frac{B'(\mu) \cdot B(e)} {B(\mu) \cdot B'(e)}\,=\, 1.00\pm0.13,
\label{rpsipdg}
\end{equation}
which is compatible with (\ref{rpsiherab}), but slightly less precise. 

Our result (\ref{rpsiherab}) for $R_{\psi'}(\mu)/R_{\psi'}(e)$
would give a significant contribution as an 
additional constraint of the global fit of $\psi'$ branching ratios
as shown in Fig.~\ref{contours}.
The error ellipse for $B'(e)$ versus $B'(\mu)$, 
evaluated by using the covariance matrix from~\cite{pdg},
is shrunk along the $B'(\mu)$ axis
(partially at the expense of a tilt, i.e. a correlation) 
after applying the additional HERA-B constraint by (\ref{rpsiherab}).

\section{\boldmath The $x_F$ dependence of nuclear suppression}      
\label{distr_adepend}      
\begin{figure}[hbtp]
\begin{center}
\epsfxsize=8.9cm
\epsfbox{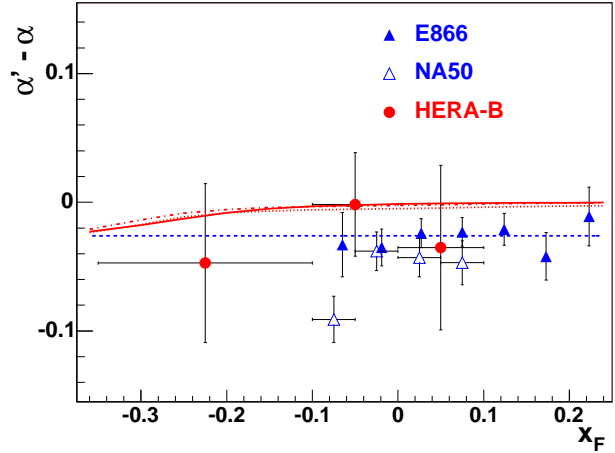}
\caption{\small
The difference of $\alpha$ 
for $\psi'$ and $J/\psi$ production 
as a function of $x_F$
determined from C and W data (circles),
compared to
results of E866~\cite{e866a} (solid triangles)  and
NA50~\cite{na50} (triangles). 
A fit of the E866 results by a constant (\ref{dae866})
is shown by the dashed line.  
The results of CEM 
for color singlet nuclear absorption~\cite{rvogt1}
at 920 GeV are displayed by the solid line.
The NRQCD results~\cite{rvogt1}
at
450 GeV and 920 GeV are shown by the dotted and dashed-dotted curves,
respectively.}
\label{dalpha}
\end{center}
\end{figure}
Usually the nuclear dependence
of cross sections is parameterized as a power law with exponent $\alpha$ 
\begin{equation} 
\sigma_A\,\,\propto\,\,{A}^\alpha,
\label{sigmaa}
\end{equation}
where $A$ is the atomic mass number. The nuclear dependence of
$R_{\psi'}$ is then also a power law in $A$:    
\begin{equation}
\frac{B'\,\sigma'_A}{B\,\sigma_A}\,\propto\,\,
{A}^{\alpha'\,-\,\alpha}
\label{ratioa}
\end{equation}
with the parameters $\alpha'$ and $\alpha$ for
the $\psi'$ and $J/\psi$, respectively.
Recent high statistics measurements of the nuclear
dependence of $J/\psi$ and $\psi'$ production for proton-nucleus collisions
were made at Fermilab E866~\cite{e866a} and at CERN NA50~\cite{na50}. 
Both results are presented in terms of 
\[\Delta\,\alpha = \alpha' - \alpha\]  
for various $x_F$ slices in Fig~\ref{dalpha}.
The two results are in reasonable agreement.
The HERA-B results for $\Delta\,\alpha$, derived from C and W data 
and also shown in Fig~\ref{dalpha}, 
are consistent with both measurements, but less accurate.
The measurements shown in Fig.~\ref{dalpha} 
indicate little or no $x_F$ dependence of $\Delta\,\alpha$. 

The calculations of nuclear suppression 
by the CEM for color singlet absorption and NRQCD models~\cite{rvogt1}
also indicate slow $x_F$ and collision energy 
dependences of $\Delta\,\alpha$ and 
that $\alpha$ is larger than $\alpha'$, 
however the predicted difference between $\alpha'$ and $\alpha$
is less than the measured difference.  
The CEM, with the additional assumptions in~\cite{kharzeev}, 
predicts that 
$\Delta\,\alpha$
should be zero over the $x_F$ 
range of this measurement.
Because of the small variation of $\Delta\,\alpha$ 
in our kinematic range~(\ref{kin_region}),
we will use the experimental value  averaged over $x_F$ 
(see Sect.~\ref{compare}) 
for the analysis of our results and for
comparison with calculations performed for $pp$ interactions. 

The E866 results, in the $x_F$ range shown 
in Fig.~\ref{dalpha}, were fitted 
by a constant value:
\begin{equation}
\Delta\,\alpha({\rm E866})\,\,=\,\,-0.026\,\pm\,0.005,
\label{dae866}
\end{equation}
which will be used as an independent measurement of 
$\Delta\,\alpha$ in the analysis of $R_{\psi'}$ results in the next Sect.

\section{\boldmath Comparison with previous measurements}     
\label{compare}    
\begin{figure}[h] 
\begin{center} 
\epsfxsize=8.9cm  
\epsfbox{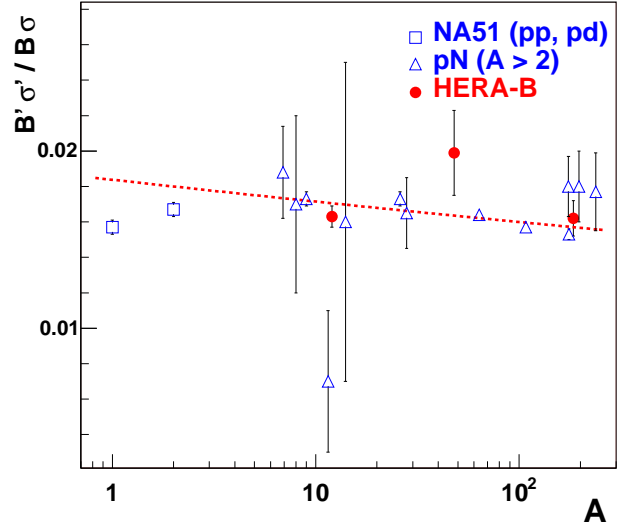} 
\caption{
A compilation of measurements of $R_{\psi'}$
as a function of the atomic mass number of the target, $A$.  
All previous measurements (Table~6) 
for $A>2$ (i.e. excluding NA51) are fitted (dashed line) 
by the function (\ref{ravsa})
with the E866 value (\ref{dae866}) used as an additional 
measurement of the power in the fit.
The HERA-B results combined for $e$ and $\mu$ 
are also displayed.}
\label{rall_vs_a_na51} 
\end{center} 
\end{figure}

A compilation of measurements of $R_{\psi'}$ is shown 
in Fig.~\ref{rall_vs_a_na51} (see also Table~6)
together with the HERA-B results for the three targets.
Previous measurements of $R_{\psi'}$ for 
atomic mass numbers $A>2$ 
were fitted 
by the function   
\begin{equation}
\frac{B'\,\sigma_A(\psi')}{B\,\sigma_A(J/\psi)}
\,\,=
\,\,R_{1\psi'}\cdot{A}^{\Delta\,\alpha},
\label{ravsa}
\end{equation}
with two parameters, where $R_{1\psi'}$ is the value of the fit 
function when formally extrapolated to $A=1$ and $\Delta\,\alpha$ is the power.
The E866 value (\ref{dae866}) was used as an additional
measurement of the power in the fit.
The fitted parameters are:
\begin{equation} 
R_{1\psi'}\,=\,0.0184\pm0.0004,\,\,\,\,
\Delta\,\alpha\,=\,-0.030\pm0.004.
\label{rall_fit_a}   
\end{equation}   
Measurements performed with hydrogen and deuterium targets do
not follow this simple power law dependence,
as seen in Fig.~\ref{rall_vs_a_na51}, where
the NA51 measurements~\cite{na51} 
for $pp$ and $pd$ interactions are also displayed.
The HERA-B results are in good agreement with the fit
of previous measurements.
\begin{table*}[hbtp]
\begin{center} 
\caption{
Ratio of $\psi'$ and $J/\psi$ cross sections in the dilepton channel
measured
in previous experiments of charmonium production
by protons with  momentum, $p$, interacting with targets with
the indicated
atomic mass number, $A$. 
The center--of--mass energy of the interacting 
nucleons, $\sqrt{s}$, is also given.
The results of NA50 are listed for the ``high intensity'' samples
since they are better suited to study the $R_{\psi'}$ ratio~\cite{na50}.
The measurements were performed in the dimuon channel,  
with the exception of the ISR experiment (last row) which detected
$e^+e^-$ pairs.}
\label{rall} 
\begin{tabular}{cccccl}  
\hline 
$p$N & $A$ & $p$(GeV/$c$)& $\sqrt{s}$(GeV)& $B'\sigma'/B\sigma$
& Experiment \\ 
\hline 
$pp$ & 1 & 450 & 29.1 & $0.0157\pm0.0004\pm0.0002$ & NA51~\cite{na51}\\ 
$pd$ & 2 & 450 & 29.1 & $0.0167\pm0.0004\pm0.00025$ & NA51~\cite{na51}\\
\hline
$p$Be & 9 & 450 & 29.1 & $0.0173\pm0.0004\pm0.0002$ & NA50~\cite{na50}\\  
$p$Al & 27 & 450 & 29.1 & $0.0173\pm0.0003\pm0.0004$ & NA50~\cite{na50}\\ 
$p$Cu & 64 & 450 & 29.1 & $0.0164\pm0.0002\pm0.0002$ & NA50~\cite{na50}\\  
$p$Ag & 108 & 450 & 29.1 & $0.0157\pm0.0002\pm0.0002$ & NA50~\cite{na50}\\  
$p$W & 184 & 450 & 29.1 & $0.0153\pm0.0003\pm0.0002$ & NA50~\cite{na50}\\  
\hline
 $p$W & 184 & 200 & 19.4 & $0.0180\pm0.0017$ & NA38 in~\cite{na38phd}\\   
 $p$U & 238 & 200 & 19.4 & $0.0177\pm0.0022$ & NA38 in~\cite{na38phd}\\  
\hline
 $p$Be & 9   & 400 & 27.4 & $0.017\pm0.005$ &       E288~~\cite{e288}\\
 $p$C  & 12  & 225 & 20.6 & $0.007 \pm 0.004$   &       E331~\cite{e331}\\
 $p$C  & 12  & 225 & 20.6 & $0.016\pm0.009$     &       E444~\cite{e444}\\
 $p$Li & 7 & 300 & 23.8 & $0.0188\pm0.0026\pm0.0005$ & E705~\cite{e705}\\    
 $p$Si & 28 & 800 & 38.8 & $0.0165\pm0.0020$ & E771~\cite{e771}\\    
 $p$Au & 197 & 800 & 38.8 & $0.018\pm0.001\pm0.002$ & E789~\cite{e789}\\   
\hline 
 $p$$p$  & 1  &  --  & 62.4 & $0.019\pm0.007$         & ISR~\cite{isr}\\
\hline 
\end{tabular} 
\end{center} 
\end{table*} 

The HERA-B measurements of $R_{\psi'}$ for three targets 
are well fitted  by the function (\ref{ravsa})
with the value $\Delta\,\alpha$ from (\ref{rall_fit_a}) 
used as an additional measurement 
in the fit,
resulting in
\begin{equation} 
\begin{tabular}{cl}
$R_{1\psi'}$ & $=\,0.0180\pm0.0006\pm0.0005,$ \\ 
$\Delta\,\alpha$ & $=-0.029\pm0.004.$  \\
\end{tabular}
\label{r1herab}   
\end{equation}   
\begin{figure}[h] 
\begin{center} 
\epsfxsize=8.9cm 
\epsfbox{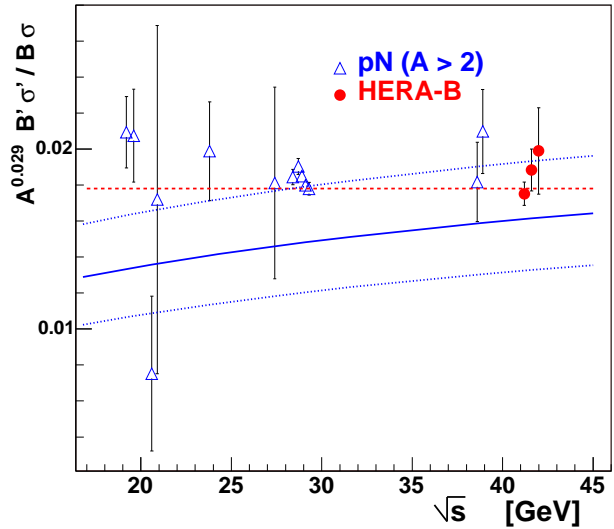} 
\caption{
A compilation of measurements of $R_{\psi'}$ in the
dilepton decay mode as a function of   
the center--of--mass energy. 
The measured ratios are rescaled by $A^{0.029}$
to compensate the nuclear suppression for targets
with differing atomic mass number $A$ 
(see Eq. (\ref{r1herab})). 
The HERA-B results combined for $e$ and $\mu$
are shown separately for C, Ti and W.
The previous measurements from Table~6
(with 
atomic mass number, 
$A>2$)  
are presented. 
The CEM expectation~\cite{rvogt2} is displayed by the dashed line. 
The NRQCD calculation~\cite{mnrqcd}  
is shown by the solid line. 
The uncertainty of the calculation is shown by dotted lines.}
\label{ra026_vs_s} 
\end{center} 
\end{figure} 
The latter $R_{1\psi'}$ value is in agreement with (\ref{rall_fit_a}). 
A joint fit of previous ($A>2$) and HERA-B measurements, 
using the value $\Delta\,\alpha$ from (\ref{dae866}) as an additional 
measurement in the fit,
results in 
$R_{1\psi'}\,=\,0.0183\pm0.0003$
and $\Delta\,\alpha$ as given in  (\ref{r1herab}). 

A compilation of measurements of $R_{\psi'}$  
as a function of center--of--mass energy
is shown in Fig.~\ref{ra026_vs_s}. 
The measurements were performed with various
targets and therefore need to be adjusted to compensate for nuclear effects.
We assume the $A$-dependence given by 
Eq. (\ref{ravsa})
with $\Delta\alpha$ from
(\ref{r1herab}) 
and therefore rescale all measurements by 
$A^{0.029}$ neglecting the error in $\Delta\alpha$.
The rescaled $R_{\psi'}$ measurements are
consistent with a flat energy dependence,
in agreement with the CEM~\cite{rvogt2}.
NRQCD calculations for $R_{\psi'}$ \cite{mnrqcd} 
show a slow increase with 
center--off-mass energy. 
Due to the uncertainties of the calculation 
such a variation of $R_{\psi'}$ 
is not excluded, but is disfavored.

\section{\boldmath Conclusions}      
\label{concl}     
We have performed a study of $\psi'$ decays into $\mu^+\mu^-$ and $e^+e^-$
using a sample of dilepton triggered data recorded during the 
HERA-B 2002--2003 running period. The sample was  divided
roughly equally
between the dimuon and dielectron decay modes. 
The analysis was based on the selection of dilepton events with relatively
low background contamination and fitting of the dilepton invariant mass
spectra in the area around the $J/\psi$ and $\psi'$ signals.
The fitted ratio of the numbers of events in the $\psi'$ and $J/\psi$ peaks
was corrected for the  $J/\psi(\psi')$ efficiency, $\epsilon(\epsilon')$, 
to obtain the ratio of the $\psi'$ and $J/\psi$ production cross sections 
in the dilepton channel, $R_{\psi'}$.

For the analysis, we selected the kinematic domain (\ref{kin_region}),
in which the $\psi'$ signal was clearly visible.
The efficiency ratio, $\epsilon/\epsilon'$, was mainly determined by geometric
factors and, therefore, was stable during the running time and 
reliable for MC calculations. 
The HERA-B results are the first measurement of $R_{\psi'}$ in 
the negative $x_F$ range. 
The measured $R_{\psi'}$  
shows little or no $x_F$ dependence
and is consistent with the previous measurements 
for positive $x_F$ and with CEM and NRQCD calculations.
The results suggest
a slow increase of $R_{\psi'}$ with increasing $p_T$,
although a flat $p_T$ dependence cannot be excluded.

From the analysis of the angular dependence of $R_{\psi'}$, 
the difference of the polarization parameters 
$\lambda$
for the $\psi'$ 
and the $J/\psi$ has been derived:
\[\Delta \lambda = \lambda' - \lambda = 0.23\pm0.17.\]  
The ratios of $\psi'$ and $J/\psi$ production cross sections in the  
dilepton channel combined for the $\mu$ and $e$ modes and
measured for carbon, titanium and tungsten targets are:
\begin{equation}  
\begin{array}{c}
B'\,\sigma'\,/\,B\,\sigma\,\,({\rm C}) = 0.0163\pm0.0006\pm0.0005,\\ 
B'\,\sigma'\,/\,B\,\sigma\,\,({\rm Ti})= 0.0199\pm0.0024\pm0.0006,\\ 
B'\,\sigma'\,/\,B\,\sigma\,\,({\rm W}) = 0.0162\pm0.0010\pm0.0005,\\
\end{array}
\label{rhb} 
\end{equation} 
respectively. 
The quoted errors are statistical and systematic.

Assuming a nuclear dependence in the form 
of a power law, $R_{1\psi'} \cdot A^{\Delta\alpha}$,
and using as a constraint  
the value $\Delta\alpha=-0.030\pm0.004$
obtained from previous measurements,
a fit to the above values yielded:
\[R_{1\psi'} = 0.0180\pm0.0006\pm0.0005.\]
With a joint fit to HERA-B results and 
previous measurements for different targets ($A>2$),
assuming a power law dependence on the target atomic mass number 
and using the value $\Delta\alpha$ 
extracted from E866 results as an additional measurement in the fit, 
we obtained the following
values for the fitted parameter:
\[ R_{1\psi'}\,=\,0.0183\pm0.0003,\,\,\,\,\Delta\alpha\,=\,-0.029\pm0.004.\]

The HERA-B results (\ref{rhb}) 
indicate 
no significant energy dependence of $R_{\psi'}$
with respect to previous measurements at lower energies.

Averaging over all three targets yields the ratio: 
\[R_{\psi'}(\mu)/R_{\psi'}(e) = 1.00\pm0.08\pm0.04\]
This result confirms $e-\mu$ universality in dilepton decays
of the $\psi'$
with an accuracy which is better than can 
be achieved using only current PDG branching ratios 
for the $\psi'$ and $J/\psi$ dilepton decays.
The result can 
be used as an additional constraint to update 
the PDG branching ratios for $\psi'$ dilepton 
decays.

\section{Acknowledgments}      
\label{acknow}     
We express our gratitude to the DESY laboratory for the strong support
in setting up and running the HERA-B experiment. 
We are also indebted to the DESY accelerator group for their continuous
efforts to provide good and stable beam conditions.
The HERA-B experiment would not have been possible without the enormous
effort and commitment of our technical and administrative staff.
It is a pleasure to thank all of them.
 
We~~thank~~R.\,Vogt~~for~~many~~useful~~discussions~~and
M.\,Beneke for advice on the treatment of charmonia polarization.


%
%

\end{document}